# OCEAN: A Five-Phase Framework for Meaningful Change

Andrea Fronzetti Colladon

*Roma Tre University, Rome, Italy*

**Abstract**

Managing meaningful change requires more than motivation, insight, or isolated techniques. This paper introduces OCEAN, a five-phase conceptual framework for structuring nonclinical developmental conversations and change-oriented work in coaching and in the practice of leaders and HR professionals. The five phases are Observe, Create, Expose, Awaken, and Nurture. In Observe, individuals develop a systemic, present-oriented understanding of the situation. In Create, they turn that understanding into goals that are meaningful and actionable. In Expose, they identify the attempted solutions that maintain the difficulty they are meant to remove, along with the resistances, biases, and self-deceptions that keep them in place. In Awaken, they activate new behaviors through planning, reframing, paradoxical movement, and emotionally significant experiences. In Nurture, they consolidate progress through repetition, habits, connection, and sustainable contexts. OCEAN integrates insights from systemic and interactional thinking, goal construction, emotional awareness, and habit consolidation into a structured process.

## 1. Conceptualizing the OCEAN Framework

Change is not an exceptional condition. It is the ordinary state of personal, relational, and organizational life. Individuals change as their needs, roles, and aspirations evolve; teams change as members learn, resist, join, or leave; organizations change because markets, technologies, and institutions change around them, and also because of internal pressures. The managerial and developmental question is not whether change will occur, but whether it will be guided or merely endured.

Yet deliberate change efforts frequently disappoint, and the shortfall is often laid at the door of leadership: leaders who can set a direction but cannot move the system behind it (Errida & Lotfi, 2021; Kotter, 1995). Such failures have structural, political, and interpersonal components, and the interpersonal ones are usually diagnosed too narrowly. Resistance, for instance, is often read as a deficit of motivation or discipline in those being asked to change, rather than as usable feedback about the change process, partly produced by the change agents themselves (Ford & Ford, 2010).

People rarely change simply by trying harder, and groups do not become more effective merely because leaders ask them to behave differently. Meaningful change requires a structured method: observing what is happening, defining goals that have motivational force, identifying the mechanisms that maintain ineffective patterns, interrupting those mechanisms through action, and consolidating what works.

I conceptualize OCEAN[1] as a framework that organizes these requirements into five phases: Observe, Create, Expose, Awaken, and Nurture (see Figure 1). The framework is intended to support nonclinical change work in coaching and in the developmental practice of leaders and HR professionals, particularly in relation to personal development, problem-solving, and performance improvement. The aim is to organize a sequence of attentional, dialogic, and practical moves that can help individuals and practitioners orient themselves in complex situations.

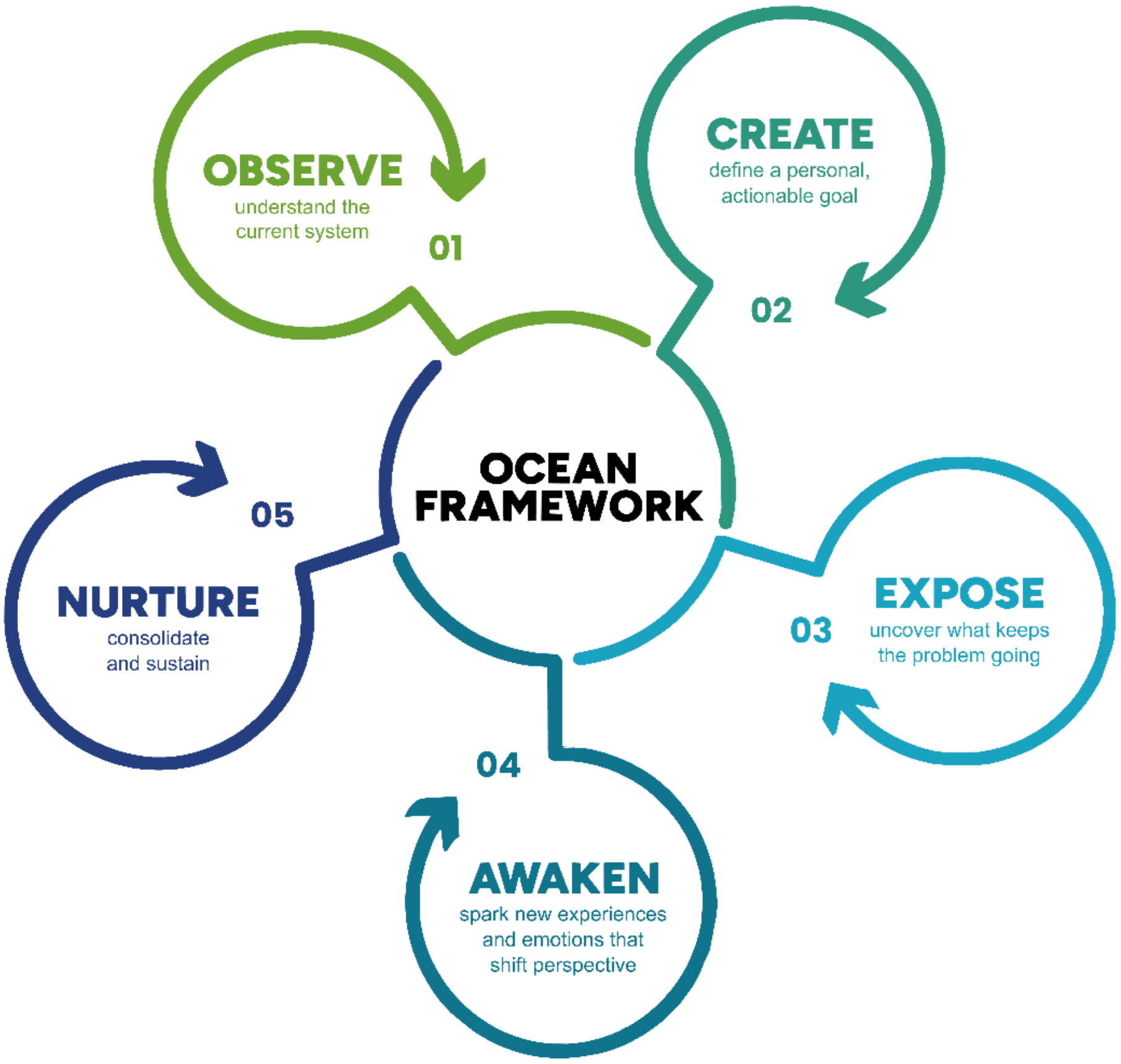

**Figure 1.** The OCEAN Framework

[1] Although OCEAN is also used as an acronym for the Big Five personality traits, the present framework uses the term independently and with an entirely different referent.

This approach integrates and reinterprets contributions from psychology, sociology, and management. It draws particularly on systemic and interactional studies associated with the Mental Research Institute tradition and related developments (e.g., Fronzetti Colladon & Grippa, 2025; Milanese & Mordazzi, 2014; Nardone & Balbi, 2017; Watzlawick et al., 1974, 2011; Watzlawick & Weakland, 1977). Accordingly, this framework treats change as a reconfiguration of meanings, perceptions, interactions, and actions, not merely as the rational selection of better behavior.

This distinction is important. In many change processes, the difficulty does not lie in the absence of information. People often know what they should do. Organizations often know that existing routines no longer work. Leaders often know that a team is stuck. The obstacle is that knowing does not automatically modify the system of perceptions, emotions, and habits that sustains the current equilibrium. Change becomes more likely when individuals and groups learn to view the present situation differently, give their efforts a precise direction, make hidden patterns visible, activate new experiences, and sustain new behaviors long enough for them to become stable without becoming rigid.

The method I outline here is not a clinical protocol, a diagnostic tool, or a substitute for medical, psychological, psychotherapeutic, or psychiatric intervention. Although it draws on psychological and systemic traditions, I position it primarily within coaching and the developmental practice of leaders and HR professionals. Accordingly, whenever change involves intense emotional suffering or otherwise calls for specialized care, referral to qualified professionals is the appropriate course.

Human behavior is shaped by complex, nonlinear systems rather than single causes. Small actions can generate disproportionate effects, while apparently strong interventions may fail because the system absorbs them without changing. The framework, therefore, attends to the content of action as well as to timing, sequence, and interactional consequences. Some changes must occur gradually, because gradual movement often reduces fear and resistance. Others begin with small actions and then accelerate through cascading effects.

The five phases form a sequence, but not a rigid linear procedure[2]. A poorly defined goal may distort the analysis of attempted solutions; a new experience may reveal that the original observation was incomplete; difficulty in consolidation may indicate that the system still rewards the old behavior. What I provide is a route that allows backward movement when new information requires revision. An application example is provided in Appendix A.

The framework speaks primarily to one literature: the process models practitioners use to structure developmental conversations. No short list can represent it, but GROW is perhaps the most widely diffused example in coaching (Whitmore, 1992), and Appreciative Inquiry (Cooperrider et al., 2008) illustrates an adjacent logic in organizational settings. Such models differ in their assumptions but tend to share a broad architecture: they establish a direction, explore the current situation, generate options, and move to action. OCEAN departs from this architecture not only in the number and sequence of its phases, but also in the specific work and tools assigned to each one. It places systemic observation before rather than after goal construction, and it adds two phases with no direct equivalent. Expose is a dedicated inquiry into the mechanisms that keep the current pattern in place, which these models ordinarily treat as obstacles met along the way. This form of inquiry is well developed in the systemic and strategic tradition (Watzlawick et al., 1974; Watzlawick & Weakland, 1977) and has been extended to coaching and problem-solving (Milanese & Mordazzi, 2014; Nardone, 2009), but it has not become a phase of the general process models discussed here. Its closest parallel elsewhere is not a process model of this kind but Immunity to Change (Kegan & Lahey, 2009), which shares the premise that the effort to improve may itself be part of what prevents improvement. Nurture treats consolidation as an active phase with mechanisms and tools of its own. Stage models of change recognize the need to sustain gains, and some explicitly include maintenance as a stage (Prochaska et al., 1992). However, their focus is on the individual's change process rather than on how maintenance is addressed within developmental conversations. In coaching, maintenance has received comparatively little attention (Theeboom et al., 2017).

[2] It can be useful, however, to define a clear and relatively short time horizon for a coaching engagement, for example around ten sessions, adjustable to the specific case. This bounded structure can itself function as a motivational lever: limited availability increases perceived value and encourages commitment (Watzlawick & Weakland, 1977).

The contribution of this paper is conceptual. Rather than replacing existing theories, the framework organizes heterogeneous insights into a process that can be used in practice. OCEAN contributes a phase architecture for change work. First, it places systemic observation before goal construction, because the first stated goal may reproduce a distorted or defensive reading of the situation. Second, it separates goal construction from the exposure of maintaining patterns, so that practitioners do not move directly from aspiration to action without examining attempted solutions, resistances, biases, and self-deceptions. Third, it links activation to both rational planning and emotionally significant experience, recognizing that insight alone often fails to change behavior. Fourth, it treats consolidation as a phase rather than a secondary follow-up because change must be stabilized through repetition, attribution, relationships, and context. The framework's contribution lies not simply in the sequence of phases, but in the interdependencies it makes explicit: initial goals may reproduce the patterns that sustain a problem; interventions may reinforce those patterns unless the problem-maintaining mechanisms are identified; and initial improvements may remain fragile when the conditions supporting previous behaviors persist. OCEAN integrates these relationships into a recursive account of developmental change, helping to explain why apparently appropriate goals and interventions may fail to produce sustainable outcomes.

It may help to distinguish three different relationships between this paper and the fields it touches. The framework contributes to process models for nonclinical developmental work. It draws on systemic and interactional thinking, goal-setting theory, research on emotion and behavior change, and work on habit formation, which supply the content of individual phases without being the target of the contribution. Where change management or leadership appear in what follows, they appear as domains of application or as sources, not as literatures to which the paper claims a theoretical contribution.

## 2. The Transformative Dialogue

Transformative Dialogue[3] is introduced here as the communicative backbone of the entire approach. It refers to the communicative process through which questions, listening, paraphrases, reframing, and emotionally resonant language help individuals observe their reality differently and move toward change. It helps the change facilitator enter the other person's perceived reality, collect details, define goals, visualize possible futures, generate commitment, reveal automatic behaviors and attempted solutions, support new actions and emotionally significant experiences, and consolidate new equilibria.

Motivational interviewing established part of the ground on which Transformative Dialogue stands: the collaborative and nonjudgmental stance, the combination of inquiry and reflective listening, and the understanding of resistance as something produced within the interaction rather than brought to it by the person (Miller & Rollnick, 2013). What follows differs in three specific respects. First, the dialogue described here changes function across a five-phase process rather than serving a single purpose. Second, it relies more heavily on questioning, where motivational interviewing privileges reflection, and it employs two families of questions that play a more limited role in motivational interviewing: alternative-response questions, which narrow the field to structured options, and orienting questions, which invite a decision. Third, it includes reframings that follow unconventional logics, in which the same situation is reformulated so that the function or cost of a behavior becomes visible, and which later in the process may be extended into agreed courses of action. Reframing is common to both, but in motivational interviewing it remains within the values and motivations the person has already voiced, whereas here it may also alter the perceived function of a behavior the person has not yet questioned.

Every exchange contributes to building and reshaping the relationship between those communicating. Watzlawick et al. (2011) emphasize the pragmatic dimension of communication: words, silence, tone, and posture do not merely transmit information; they influence responses and behaviors. A Transformative Dialogue, therefore, begins in a receptive stance. The starting point is relational attunement. People are more willing to open up when they feel heard, understood, and not judged. Perceived similarity, connection, and genuine interest increase openness and collaboration (Cialdini, 2006). Similarly, leaders often build trust not by appearing infallible, but by communicating transparently and acknowledging limits when appropriate (Brown, 2018).

Accordingly, the coach or leader usually starts from a *one-down* position, asking and listening so that the other person's world can be revealed (Haley, 1963; Watzlawick et al., 2011). Later, especially when concrete actions are defined, the facilitator may take on a more active role. Moreover, attunement is built by calibrating the content, pace, and depth of the dialogue to the person and the setting, so that the facilitator adopts a complementary role and avoids simply mirroring the other party[4].

[3] The term exists in other theoretical and practical traditions. Here, I use it in a specific sense, independently of those previous uses. It partly overlaps with Strategic Dialogue, especially in the use of paraphrases, reframing, and progressive convergence toward action (Nardone & Salvini, 2018). It differs, however, in the stronger use, particularly at the beginning, of open questions, knowledge-building questions, probing questions, and deep questions.

Questions are the main instrument because they collect information while also guiding attention and influencing what the interlocutor becomes able to see and reconsider. A question can open alternative perspectives, reveal exceptions, clarify meanings, and make possible solutions more visible. Questions are among the primary tools used to encourage self-reflection and facilitate change (Moos & Spranz-Fogasy, 2024). Premature interpretation is a common error, especially when the facilitator recognizes a familiar pattern and assumes that the present case is identical to previous ones. Intuitions can be valuable, but they should be suspended and tested against what the interlocutor actually says, in line with the need to delay premature conclusions when evidence is incomplete (Kahneman et al., 2022). Questions that carry an implicit judgment or already contain the answer should be avoided because they invite defensiveness rather than reflection.

The Transformative Dialogue uses seven families of questions:

- *Open and introductory questions* create the initial space for the dialogue, allow the person to introduce the issue in their own language, and can foster rapport and openness (Simpson & McDowell, 2019). For example: "What brings you here today?"
- *Knowledge-building questions* broaden the context by collecting information about actors, situations, timing, relationships, and constraints. For example: "Who is involved when this difficulty emerges, and when does it usually happen?"
- *Probing questions* reveal hidden details and convert general claims into observable mechanisms. They are useful when the person uses broad expressions such as "nothing works", "the team is blocked", or "I am not confident". The question helps transform these statements into concrete episodes, behaviors, and sequences. For example: "When you say that the team is blocked, what exactly happens during the meeting?"
- *Deep questions* touch emotions, values, identity, and meaning. Instead of asking "How did the interview go?", one might ask "Was there a moment when you realized you really wanted to work for that company?" Questions like these can foster deeper attunement and trust. Consistent with this idea, Aron et al. (1997) found that structured sequences of personal questions can promote perceived interpersonal closeness, while Ludwig et al. (2022) argued that sharing emotions, whether positive or negative, can increase feelings of closeness, thereby encouraging further sharing.
- *Alternative-response questions* narrow the field after exploration and help build a funnel toward understanding and action (Nardone & Salvini, 2018). They offer structured alternatives that allow the facilitator and the interlocutor to test hypotheses without imposing an interpretation. For example: "Do you tend to react by avoiding the situation, or by trying to control it more?"

[4] For example, broader, narrative, open-ended questions work better with taciturn people who need space to elaborate, while shorter, closed, alternative-response questions help focus interlocutors who tend to ramble without reaching a point.

- *Reframing questions* introduce a different perspective. They help the person look at the same situation from another angle and notice the possible function or cost of a behavior. The aim is not to tell the person that their view is wrong, but to make a more useful reading available. For example: "If your team interpreted your constant availability not as support, but as a sign that they cannot decide alone, what would change in the way you read the situation?"
- *Orienting questions* invite a choice that points toward action. They usually appear after sufficient exploration, paraphrasing, and reframing, when the person is ready to decide how to proceed. They connect the dialogue to a concrete next step while preserving ownership. For example: "Given what we have discussed so far, would it be more useful for you to continue asking for help as you have been doing, or to temporarily stop asking for support and observe what happens?"

*Active listening* gives these questions their effect. It means listening not only to the words being said, but also to the meanings, emotions, needs, and implicit signals behind them. In this sense, a good question is not simply deeper in content but better attuned to what the other person needs from the conversation. Indeed, active listening also requires sensitivity to the function of the exchange. Conversations often serve different purposes: some are practical and oriented toward problem-solving, others are emotional and seek validation or support, and others are relational, helping people feel recognized and connected. When the response does not match the purpose of the exchange, support can be less effective and may create frustration (Duhigg, 2024; Feng, 2009; Jones, 2011; Sullivan et al., 2010). For example, if someone says they had a difficult interview, immediately asking, "What will you do differently next time?" may be useful in a problem-solving conversation. Still, it may feel premature if the person's first need is emotional support. Attending to the other person's emotions also requires the facilitator to regulate their own.

*Paraphrasing* is used to check understanding, show attention, and organize what has emerged in the interaction. It also helps the facilitator assess whether the dialogue is moving in a productive direction by verifying that the meanings being returned correspond to the other person's experience. It does not merely reproduce what has been said. It selectively returns the most relevant meanings, helping to clarify the structure of the issue under discussion (Nardone & Salvini, 2018). *Repetition* of concepts further reinforces this process by directing attention toward the elements that appear most significant. When a shared insight, a possible direction, or a relevant distinction emerges, restating it in varied formulations helps stabilize it, make it more salient, and support its retention in memory. *Reframing* then goes one step further: it reformulates facts, emotions, or interpretations so that the same situation can be viewed from another angle. The strongest form of reframing relies almost exclusively on the person's own words but emphasizes different elements or relationships among them, allowing alternative interpretations or responses to emerge and bringing previously overlooked perspectives into view (Luciani & Convertini, 2024; Watzlawick et al., 1974).

The dialogue favors discovery over instruction. When the next step emerges from the person's own answers, the person experiences authorship. Pascal's (1671) observation that people are more persuaded by reasons they discover for themselves than by reasons suggested by others captures this mechanism. Advice often fails because it comes from outside the person's construction of reality. By incorporating paraphrasing, a well-guided dialogue allows the person to reach a conclusion that feels self-generated and coherent with their view of reality.

A skilled facilitator must also attend to the emotional register of the exchange. A purely rational explanation or instructive language is often insufficient because many blocks are emotional, relational, or perceptual. The use of stories, metaphors, analogies, images, tone, rhythm, and pauses can make a new perspective vivid. Storytelling is powerful because people think and remember through narratives (Schank & Abelson, 1995), and vivid stories can activate brain regions associated with experiencing the events described (Speer et al., 2009). They matter because transformation often requires an experience of meaning that the person can remember and carry into action.

Transformative Dialogue is a method for communication, development, and problem-solving, not a clinical intervention. Its quality does not lie in flawless performance, but in sustained attention, ethical intent, and the capacity to recalibrate. It transforms observation into a shared interpretation of the situation, allowing change to begin from meanings and possibilities that the individual recognizes as their own rather than from explanations imposed by the facilitator.

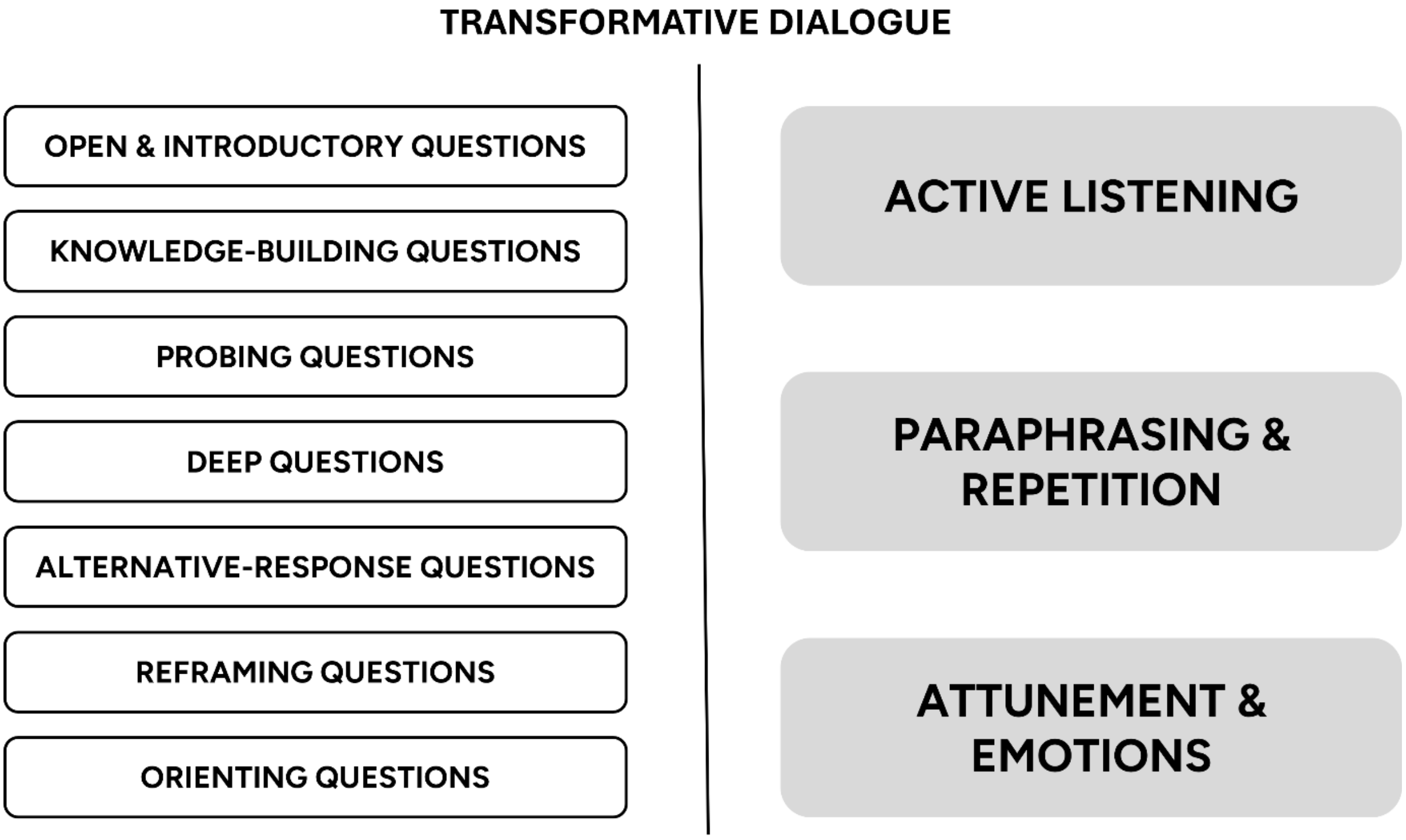

**Figure 2.** Key Elements of the Transformative Dialogue

### 3. Observe

Change begins with a disciplined reading of the present. Careful observation means attending to details, interactions, contexts, implicit rules, and subjective interpretations before moving toward explanation or intervention. In coaching and leadership work, this is often the difference between acting quickly and acting effectively.

Before defining goals, exposing attempted solutions, or activating new behaviors, one must learn to *see more*. Observation is active and requires time, attention, and restraint. Many errors in problem-solving come from excessive certainty rather than ignorance. People often act on what they think they already know, while the most relevant signals remain outside their field of attention.

The same scene can support different interpretations depending on what the observer notices and where attention is directed. The point is not that every interpretation is equally useful, but that a single interpretation is rarely sufficient. The framework, therefore, treats reality as something experienced through filters, rather than as something true or false in absolute terms. Von Glasersfeld (1984) described truth as "viable": a representation is useful when it helps individuals orient themselves and act coherently within their context. Accordingly, a person's perceived reality matters because it produces real emotions and behaviors. A message intended as neutral may be perceived as hostile; silence may be read as disapproval; a market signal may appear as either threat or opportunity.

The observer's first task is not to impose a more correct representation, but to understand how another actor constructs the situation. A manager who sees a team member as resistant may ask why the person refuses change. Better observational questions would ask what risks they perceive, what emotions they experience, and what prior interactions make resistance appear reasonable. Similarly, in consumer understanding, the analyst should not only ask what a customer buys, but what the object means, which identity it supports, and which social codes make it desirable (Roberti, 2021).

Presence is required for this kind of observation. It includes attention to verbal content, paraverbal signals, posture, gestures, gaze, silence, timing, spatial arrangements, and context. Selective attention allows people to concentrate on relevant stimuli, but it can also become a source of blindness (Kahneman, 1973). The observer must focus enough to reduce noise while remaining open to unexpected signals. Goleman (2014) distinguishes internal focus, focus on others, and external focus. The distinction matters because observation fails when a manager understands the market but not the team, or the client's stated narrative but not the emotions beneath it. In general, the quality of attention affects the quality of observation. This also matters in organizations, where pressure, time scarcity, and emotional activation often narrow perception (Easterbrook, 1959; Galván & Rahdar, 2013; Karau & Kelly, 1992).

A second concern is *systemic thinking*. Individuals communicate continuously with themselves and with others. Watzlawick et al. (2011) remind us that one "cannot not communicate": silence and absence are still messages, and every message conveys, besides its content, how those communicating define their relationship. Human systems are organized by explicit rules, such as procedures and roles, and by implicit rules, such as who may disagree, which topics are avoided, and how mistakes are treated. These rules influence what members notice, ignore, or normalize.

This view is consistent with the idea that facts cannot be fully understood outside the relational context in which they appear (Watzlawick & Beavin, 1967). Systems also tend toward *homeostasis*. Once a pattern is established, the system activates mechanisms that restore it after disruption. Jackson (1965) showed how repeated interactions create relational rules that resemble implicit contracts, while Haley (1977) illustrated how problems may reside in relational configurations rather than individual traits. In organizations, a manager seen as indecisive may in fact be caught between two informal coalitions with opposing demands, and replacing the manager would leave the problem in place. Because homeostatic systems absorb changes made within their rules (first-order change), durable change often requires changing the rules themselves (second-order change) (Watzlawick et al., 1974). One route is *metacommunication* – communicating about communication – which makes those rules discussable, often with the help of an external observer such as a coach.

This systemic orientation also argues against an excessive reliance on *why* questions. "Why" can be useful, but it often directs attention toward causes and blame. A more generative question is "How does this work?". Ashby's (1956, p. 11) point that "transformation is concerned with what happens, not with why it happens" is aligned with this functional orientation. For example, questions about *what* happens, *who* participates, *where* the process loses momentum, *when* the difficulty arises, and *how* the current system makes the result more likely shift attention from blame to mechanism.

Individuals can also learn to observe themselves from the outside, for instance by describing a recurring situation in the third person, including their own posture, words, and the reactions of others, without judging or seeking solutions. Digital tools may support the observation process by functioning as mirrors. Many signals in contemporary work are nonverbal or paraverbal, such as email response time, rhythm of communication, emotional tone, and turn-taking. Pentland (2008) described such signals as "honest" because they are less controlled than explicit behavior; they are subtle patterns in how people interact. AI and analytics systems can make these patterns visible, offering feedback that helps individuals observe themselves from the outside (Fronzetti Colladon & Grippa, 2025; Fronzetti Colladon & Vestrelli, 2025; Gloor et al., 2017). These modern tools should provoke inquiry, not surveillance or reductive judgment.

Finally, the facilitator should help individuals extend their observation beyond immediate relationships to the wider social and cultural contexts in which choices are made. In environments characterized by abundant options and shifting identities, seemingly individual decisions are shaped by social meanings, significant bonds, and the way alternatives are presented (Bauman, 2000; Dahrendorf, 1979; Thaler & Sunstein, 2008). Attending to these influences brings into view the contextual rules that shape behavior.

### 3.1. Present-Oriented Inquiry in Change Processes

Observation of the present is privileged because it is the moment in which behaviors can be directly observed, tested, and guided. This does not mean that the past has no relevance. Past experiences, learned scripts, and prior relationships can influence current behavior. However, the past becomes useful mainly when it manifests itself in present interactions, perceptions, and attempted solutions (Watzlawick & Weakland, 1977). The central object is not the reconstruction of what once happened, but the identification of what is happening now and how current patterns can be modified.

This position responds to a common but problematic assumption: that when a problem exists, one must first find its origin. Such inquiry may create the impression of depth, but it can also move attention away from the mechanisms that maintain the problem today. The past cannot be observed directly; it can only be remembered, narrated, and interpreted from the standpoint of the present. Memory is selective, and what people remember is shaped by current mood and emotional state (Faul & LaBar, 2023). The present, by contrast, can be observed in action, and the effects of small changes can be assessed in the current system.

A present-oriented approach also reduces the risks of blame and labeling by shifting attention from individual faults to ongoing interactional patterns. Linear explanations tend to isolate a single actor or cause, whereas many problems arise from *circular* processes in which each behavior both responds to and reinforces the other (Watzlawick et al., 1974). For instance, a manager may become more controlling because the team appears passive, while the team may become more passive because the manager controls every decision. From this perspective, the problem does not reside solely in an "authoritarian manager" or an "unmotivated team", but in the self-reinforcing loop between them. It follows that not everyone must change at once: a change by one actor can alter the loop and, through it, the responses of the others involved (Fisch et al., 1982).

Labels such as "resistant", "fragile", or "difficult" can become "self-fulfilling" and obscure the conditions under which behavior emerges. Once a person or group is described in this way, others may begin to interpret their actions through the label and respond accordingly, reinforcing the very behavior they expect to see. Watzlawick and Weakland (1977) also showed that behavior that appears problematic at the individual level may serve a function within a wider system. For example, an employee labeled as resistant may actually be protecting another team member they care about by expressing doubts the other person shares but does not have the courage to voice.

The past should therefore be examined only to the extent that it is active now. If an old pattern shapes current distrust, it matters because it affects present interpretations, conversations, and decisions. The relevant question is how past learning organizes current behavior and how that organization can be changed. This stance also clarifies why "insight" alone cannot be the main criterion for success in a change process. Insight may help, but the practical questions are whether the problem has been reduced, the goal has been reached, and the person, team, or organization acts more effectively. Watzlawick et al. (1974) warned against approaches in which the absence of insight is treated as a problem, even when the original difficulty has improved. Fisch (1965) likewise criticized the tendency to dismiss rapid improvement as superficial. Early progress can strengthen motivation, self-efficacy, and willingness to continue, provided that it is later consolidated.

The present, then, is the methodological center of the OCEAN framework: the place where interactions can be observed, complex systems mapped, attempted solutions interrupted, goals tested, and results measured.

## 4. Create

During the Create phase, observation becomes direction. After the interaction system has been explored, the central task is to define a future state that is clear, meaningful, and actionable. The phase connects the visualization of future scenarios to concrete execution, drawing on resources that are often already present in the individual or organization. Here, actors are required to clarify what should change, why it matters, and how progress can be recognized and measured.

Many change efforts fail before action begins because the objective is vague, imposed from outside, detached from the individual's values, unrealistic, unmeasurable, or based on an idealized image of what life or work should be. A team may be asked to become more innovative without clarifying what kind of innovation is expected or what trade-offs the organization will accept. An individual may want to be more confident or productive without translating that aspiration into observable behaviors. This phase addresses such ambiguity.

The first requirement is *ownership*. A goal should feel like one's own, or at least be negotiated and shared in a way that generates commitment. Goals imposed by family, peers, managers, organizational culture, or social expectations may produce compliance but often lack motivational depth. The question is not only what the individual wants to do, but who they want to become and whether the goal is coherent with identity, values, and long-term aspirations. Otherwise, change is likely to remain superficial, externally imposed, and thus fragile and prone to reversal. Ownership, however, does not imply isolation. Individuals are embedded in relationships, and organizations operate within stakeholder systems. For example, a sales target may energize a team in the short term, but if it is pursued by means of increasing pressure, ignoring workload, or weakening trust, it can undermine motivation, retention, and collaboration. In this case, the apparent performance gain undermines the conditions that enable sustainable performance.

When several problems or ambitions coexist, goal prioritization is necessary. Progress in one area may create positive effects elsewhere, but without a clear priority, the change process can become too broad and unfocused. One should also distinguish goals that can be acted upon, goals that require negotiation, conditions that require acceptance (something that cannot be changed), and expectations that should be released. The distinction protects change processes from two errors: trying to control what cannot be controlled, and pursuing objectives that do not truly matter to the person or group involved.

Goal definition should combine specificity with personal commitment or autonomous ownership. Goal-setting theory posits that specific, challenging goals enhance motivation and performance, especially when people are committed to them and receive feedback on progress (Erez, 1977; Locke, 1968). Related research on self-concordance and autonomous motivation has shown that goals are more effective when they are personally endorsed and aligned with internal motivation (Judge et al., 2005; Sheldon, 2014). Vague goal formulations such as “improving communication” or “becoming a better leader” do not provide enough direction. The SMART framework, originally introduced by Doran (1981) and subsequently adapted in practice, remains useful as a heuristic for defining objectives that are specific, measurable, achievable[5], relevant, and time-bound. The OCEAN framework aligns with this logic but adds relational and motivational conditions: the objective should be defined through dialogue with the person or group who must pursue it. Also, it must be agreed upon. A technically well-formulated objective can still fail if it is not felt as legitimate or shared.

Measurement is part of the work. Objective indicators may include sales, delivery times, errors, absenteeism, or customer satisfaction. Qualitative outcomes, on the other hand, can be tracked through self-evaluation scales, particularly in coaching. A simple rating from 1 to 10 can help individuals or teams assess where they are now, where they were before, and what would indicate a small improvement. This turns subjective experience into a shared reference point for reflection and action. At times, the scale’s usefulness can be increased by defining selected anchor points with concrete, observable descriptions. It does not replace rigorous evaluation, but can create a disciplined habit of noticing improvement.

Visualization can make the goal clearer and more motivating. The “miracle question”, particularly valuable in this respect, asks the person or team to imagine that the problem has been solved and to describe what would change, how they would notice it, and what they would do differently (De Shazer, 1988). Its value lies in shifting attention from the problem to a desired reality and in revealing behavioral and relational indicators of success.

Indeed, expected realities can influence present behavior. Merton’s (1948) concept of the *self-fulfilling prophecy* suggests that expectations about a future state can shape present behavior in ways that help bring that state about. Mental scenarios can also generate bodily and emotional activation similar to direct experience (Williams et al., 2017), and deliberate imagery has long been used to support goal-directed behavior and self-efficacy (e.g., Renner et al., 2019; Simonsmeier et al., 2021). The practical implication is that a clear and emotionally credible future makes aligned behavior easier to identify and perform, provided the visualization remains connected to present obstacles and required effort, since positive fantasies about a desired outcome, as distinct from realistic expectations of success, can reduce the energy invested in pursuing it (Oettingen & Mayer, 2002). The technique of acting “as if” extends this logic: a person performs small behaviors today that are consistent with who they are trying to become (Watts et al., 2005).

One should also avoid utopian goals and oversimplification. The search for perfect solutions can become a problem in itself (Watzlawick & Weakland, 1977). A leader may reject incremental solutions because, in their view, transformation should solve everything at once; a worker may reject every job because the ideal job should contain no frustration. Instead, meaningful objectives should be ambitious but compatible with the complexity and imperfection of real systems. When an action fails, the result is best treated as *information* and as an opportunity to learn and better understand the mechanisms behind a problem.

---

[5] Whether a goal is “easy” or “difficult” is not an inherent property of the goal itself but depends on the resources, skills, and time available to the specific actor; the same task can be trivial for one person and impossible for another.

The phase is complete when the actor has a clear and shared vision, an actionable objective that is challenging yet achievable, and indicators of progress. A preliminary idea of a first step that can be taken without waiting for perfect readiness may also emerge. The goal defined here is the reference against which Expose identifies attempted solutions, and Expose may show that it needs to be revised.

## 5. Expose

During the Expose phase, individuals bring to light the mechanisms that keep them trapped in ineffective behavioral patterns. After the current system has been clarified through observation and a meaningful objective has been defined, attention turns to identifying what currently sustains the gap between the present and desired state. In many cases, the answer does not lie in the original cause of the problem, but in *repeated attempts* to solve it. These attempted solutions may appear reasonable, even necessary, yet they maintain or aggravate the difficulty they are meant to remove (Watzlawick et al., 1974).

Trying to solve a problem is not inherently problematic. However, a difficulty may become a fully consolidated problem when individuals continue to apply the same strategy after it has ceased to be effective or has begun to produce harmful secondary effects. The analysis is not meant to assign blame; rather, it should carefully examine the mechanisms that sustain the problem. Problems often persist because actions, reactions, personal interpretations, and routines become organized into self-reinforcing loops. Individuals frequently continue doing something that aggravates a problem simply because they perceive no alternatives, or because the repeated strategy appears to be the most logical course of action. Watzlawick et al. (1974) described the paradoxical effect of doing more of the same: the stronger the effort in the wrong direction, the farther the person moves from the solution. Trying to force spontaneity, trying not to think of something, or trying to control every sign of uncertainty may preserve the very difficulty one wants to reduce.

Attempted solutions work on three levels: *thought*, *communication*, and *action*. At the level of thought, for example, a person may repeatedly tell themselves that they are not capable, thereby activating a negative self-fulfilling prophecy. At the level of communication, someone may avoid expressing disagreement to prevent conflict, allowing resentment to accumulate. At the level of action, a leader may respond to poor performance by increasing control, thereby reducing initiative and learning. These solutions become counterproductive when individuals act when inaction would be preferable, fail to act when action is required, or act in the wrong way (Fronzetti Colladon & Grippa, 2025). Inaction can itself be a solution attempt when avoidance temporarily reduces discomfort but maintains the larger problem.

This analysis requires a high degree of specificity. Habits are central to the inquiry. When behaviors repeat, they become easier to reproduce and harder to question. Repetition also reduces observation because people filter out signals that contradict their choices. Some guiding questions are: What has already been tried? Who did what? In which context? Was it conscious or automatic? What happened immediately afterward and over time? Did the action reduce the difficulty, shift it elsewhere, amplify it, or create new indirect costs? What can no longer be done, stopped, or done differently because of the problem? These questions preserve the framework's present-oriented logic while allowing selective use of the recent past. Indeed, complex systems are often *path-dependent*: a reaction that seems excessive today may be understandable when one considers the earlier events that led to it. For example, a manager's irritation at a minor delay may appear disproportionate if observed in isolation, but it becomes more understandable when the same delay has recurred despite previous clarifications and adjustments. In this regard, the useful past is the past that still appears in the present: the events that continue to shape how individuals perceive and react today. The point is not to reconstruct an entire biography or organizational history, but to identify the recurring behaviors that still affect the current pattern.

Ineffective attempted solutions often restrict the action space. Individuals may no longer be able to imagine alternatives, may know what could work but be unable to implement it, or may implement a new solution but fail to sustain it because of emotional, relational, or systemic consequences (Milanese & Mordazzi, 2014).

The first practical goal of this phase is *awareness*. Once a person recognizes that a repeated solution feeds the problem, that solution becomes less neutral. Awareness begins to restructure perception. The second goal is *interruption*. However, refraining from action can sometimes be mistaken for doing the opposite, and that is not always conducive to sustainable change. A person who overworks may try to disengage completely, or a manager who controls too much may withdraw guidance altogether. Instead, many useful solutions lie in the middle range: a more sustainable rhythm, a more flexible rule, or a negotiated boundary.

*Past exceptions* are also worth exploring, as they may point to solutions that can be reused. Sometimes the person, team, or organization has already solved the problem temporarily or performed better under specific conditions. These exceptions show that resources may already exist, but they must be tested for current functioning and replicability. A solution that worked with a previous manager or in a previous market may fail in the present.

The *worsening question* is a useful tool for bringing attempted solutions to light. It invites individuals to consider what would deliberately maintain or exaggerate the very behavior that keeps a problem in place rather than suppress it (Fisch et al., 1982; Watzlawick et al., 1974). Nardone (2009) formalized this logic into a specific analytic question: "What could you do, say, or think to make the problem worse on purpose?" It can reduce the performance pressure associated with finding the "right" solution and support the analysis of concrete, verifiable behaviors rather than abstract intentions, making avoided or automatic patterns easier to identify. An alternative technique is *problem reversal*: the actor imagines a problem opposite to the one at hand, looks for solutions, and explores that scenario in detail (e.g., De Bono, 1970). Behaviors that would solve the opposite problem often turn out to be attempted solutions that maintain the actual one. Both techniques draw on a paradoxical logic, encouraging the mind to move against its habitual response. When a person is asked to consider how to deliberately worsen the problem they want to solve, taken-for-granted behaviors become more visible, as do new possible behavioral patterns.

Resistance to change should also be analyzed in this phase. People may resist because the proposed change conflicts with their goals, is imposed by others, comes from someone they do not trust or like, seems too demanding, threatens identity, activates overly strong emotions, or would remove secondary gains. This reinforces the importance of goal ownership. A change initiative must connect with the goals, constraints, and perceived reality of the people expected to enact it. Otherwise, it may generate compliance in discourse and resistance in practice.

Different forms of resistance require different responses. Some individuals oppose openly. Others oppose indirectly or appear collaborative while acting against the change. Others want to collaborate but lack practical or emotional resources. Still others hold rigid interpretations that do not yet allow alternatives (Watzlawick & Nardone, 1997). Depending on the resistance in place, different intervention strategies should follow. For example, resistance to an uncomfortable message can sometimes be reduced by anticipating it verbally[6] before delivering it, thereby lowering the likelihood of immediate rejection or defensive closure. With apparent collaborators, small concrete tests can verify collaboration. With emotional blocks, a rational explanation may not be enough. With rigid beliefs, direct contradiction often strengthens resistance, whereas small doubts, indirect questions, and reframing may loosen it. With those who oppose change, explicitly inviting resistance can be a strategy for neutralizing their oppositional intent.

When someone says, “nothing can be done”, this often signals resistance. Yet directly opposing the claim may reinforce defensiveness. A paradoxical move may either temporarily accept the impossibility, even appearing more pessimistic than the person so that they begin to argue for change, or ask what should be done to keep the situation unchanged or *make it worse*. Nardone (2003, p. 65) captures this logic as “killing the snake with its own venom”. The same principle resonates with the idea that the best victory is achieved without fighting.

Indirect forms of communication can reduce resistance. Negation can indirectly introduce a concept, making it cognitively present and influencing judgment without explicit assertion (Autry & Duarte, 2021; Maciuszek & Polczyk, 2017; Nardone, 2014a). In the clinical and experimental literature, confusion, used ethically and cautiously, can interrupt rigid patterns and make subsequent clarification more acceptable[7] (Dolinski & Nawrat, 1998; Erickson, 1964; Watzlawick, 1993).

Bias and self-deception are also addressed in this phase. Bach (1981) described *self-deception* as a motivated process by which individuals avoid sustained or recurrent thoughts that conflict with what they wish were true. For example, people may deny a problem, oversimplify it, rationalize[8], or minimize inconvenient evidence (Freud, 1992; Goleman, 1997; Watzlawick et al., 1974). Research has also shown that automatic judgments and heuristics frequently guide repeated behavior (e.g., Cialdini, 2006; Kahneman, 2013). Individuals are subject to biases, including overgeneralization from limited experience, anchoring on initial information (Tversky & Kahneman, 1974), and attribute substitution (Kahneman & Frederick, 2002). Biases can also become collective. Groupthink suppresses dissent and protects harmony at the expense of decision quality (Janis, 1972), while polarization and cascade effects can be amplified by information environments that confirm existing beliefs (Sunstein, 2021). Metacognition represents a key corrective: individuals should ask what they are assuming, what they take for granted, which evidence would challenge their view, and what alternative explanations could produce the same signals.

[6] E.g.: “I am about to say something you might not like”.

[7] These mechanisms are reported here to explain how rigid patterns loosen, not as procedures for nonclinical practitioners. Deliberately inducing confusion requires clinical training and safeguards, and it is incompatible with the transparency that a coaching, leadership, or HR relationship requires.

The phase is complete when the actor has identified recurring attempted solutions, the levels at which they operate, what sustains them, the biases that protect them, the resistances that may block change, any useful exceptions, and the first patterns that need to be interrupted.

## 6. Awaken

Awaken moves from clarification to activation. Understanding a pattern is necessary but rarely sufficient. People may know exactly what they should stop doing and still do it. Teams recognize a routine as counterproductive and reproduce it anyway. Leaders may understand that micromanagement erodes autonomy, yet exert more control as pressure increases. In this phase, actors should focus on experiences, actions, and reframings that interrupt established patterns and make new possibilities emotionally and behaviorally available.

Two distinct pathways are available. The first is more rational and instructive. It is appropriate when the emotional load is limited, when the individual has sufficient resources to act, and when the problem responds to planning, learning, and execution. The second is experiential and often counterintuitive. It is necessary when the individual is held back by emotions, rigid beliefs, or repeated solution attempts that explanation alone cannot interrupt. The two logics typically intersect: a rational plan may require emotional experiments to become viable, while an emotional shift may need structure to hold.

When rational change is possible, the first task is to expand the solution space. People often stop at the first plausible solution because the mind economizes effort. Creativity tools such as brainstorming, mind maps, and the Six Thinking Hats technique help people approach the same problem from different positions (De Bono, 1985). The value of these tools lies in disciplined movement across perspectives. A person who habitually emphasizes risk may need to examine opportunity; a team that always starts with enthusiasm may need to examine constraints. However, these tools are often insufficient on their own, because identifying a better option is not the same as being able to implement it.

[8] Rationalization here refers to the defense mechanism by which a person creates apparently logical explanations to justify feelings, actions, or failures that may be driven by less acceptable motives.

Rational change can often benefit from decomposition. *Backward planning* is a technique that begins by analyzing and visualizing a desired final state and then proceeds by reconstructing the steps that must precede it by moving backward (Wiese et al., 2016). By moving from the endpoint to the present, the individual can identify intermediate conditions, anticipate obstacles, and define the next smallest action to take. This method has been shown to be associated with greater motivation, reduced perceived time pressure, and improved goal-relevant performance (Park et al., 2017). It can turn a difficult goal into smaller, attainable subgoals. Moving ahead in small, more easily accomplished steps can increase commitment and self-efficacy. However, very detailed subgoals may also reduce flexibility and self-organization (Rai et al., 2023). For this reason, planning should include periodic recalibration.

Nonetheless, many problems cannot be solved by planning alone. Counterproductive patterns often survive rational disconfirmation because they are sustained by emotions. A person may know that a presentation is unlikely to be disastrous and still feel apprehensive. A manager may know that delegating is necessary and still feel exposed. In such cases, the individual must first undergo a different (emotional) experience; only afterward can the mind reorganize the meaning of the situation. This can happen, for example, when an individual expects one outcome but experiences another, so that a consolidated perception begins to change (Alexander & French, 1946).

Transformative Dialogue is a key route into this process because questions, paraphrases, reframing, and emotionally resonant language help the individual reinterpret a behavior and test new responses. Reframing is not meant to discover an absolute truth hidden behind problems. It becomes valuable as soon as it leads to viable new actions, opens a behavior that was previously unavailable, and interrupts old, counterproductive patterns.

In addition, reasoning through unconventional logics can help individuals move beyond ordinary explanation, create new experiences, and open perspectives unavailable within their current frame. Paradox is one of these logics. In communication, a paradox occurs when incompatible messages coexist across levels – for example, when a manager says, “I want your honest opinion”, while their tone and facial expression signal that criticism is unwelcome. Double binds are a particularly problematic example, with blocked metacommunication and contradictory messages occurring in significant relationships (Bateson, 1972; Watzlawick & Weakland, 1977). The relevance here is pragmatic: paradoxes do not only describe communicative traps; they can also suggest a way to act on the logic that maintains the problem. A logical shift that moves paradoxes from communication to action occurs when automatic behavior is made a deliberate, observable task.

A paradoxical move may therefore consist of asking the person to *do more* of the very behavior that has been keeping them away from the desired change, but under clear and limited conditions (low-risk, never applied to a symptom or to clinically significant distress). This changes the status of the behavior. What was previously experienced as something that simply happened becomes something that can be chosen, delimited, and eventually interrupted. When a behavior is no longer merely reactive, but voluntary, its effects become manifest: the individual can observe the mechanism that fuels the problem, transform something seemingly uncontrollable into something controllable, and change the meaning of the behavior itself. When an attempted solution is scheduled, it often loses its original function. For instance, scheduled control may become less reassuring or, by being delimited, may lose part of its automatic pull. Similarly, when opposition is deliberately invited rather than resisted, it may lose its oppositional function and open the way to cooperation. Experiments can also introduce variation: a behavior is maintained, but with different timing, methods, or limits, thereby effectively changing the system's rules.

If perfectionism delays action, a small planned imperfection can loosen the rule that only flawless work is acceptable. If repeated checking reduces worry in the short term, agreeing on a controlled and amplified repetition can make checking less attractive and more conscious. If a team endlessly reopens the same discussion to avoid the risk of making a decision, dedicating a fixed, explicit time to generating all possible objections can make the pattern visible and help the group move toward a decision. In each case, the aim is not to reinforce the old pattern, but to interrupt its automaticity. The same logic that maintained the problem can become the lever through which the system begins to change (Watzlawick et al., 1974, 2011). Restraining moves apply the same logic when they discourage change with the aim of fostering it: advising the person to go slow, voicing doubts about the risks of improvement, or, once progress appears, proposing a brief, deliberate return to the old behavior (Fisch et al., 1982).

Any paradoxical experiment should meet strict conditions. It should be based on careful observation, connect to the person's own goal, and remain small, concrete, reversible, and ethically appropriate. It should be followed by a reflection on what happened. It should not manipulate or impose control. Such strategies are delicate and belong only in low-risk situations with people who have the rational and emotional resources to tolerate the experiment. Intense, persistent, generalized, or impairing distress requires stopping the work and referring the person to a qualified professional. Experiments of this kind are proposed and agreed upon, never prescribed, and the person can decline them or stop them at any point[9].

While a paradox presents incompatible messages at the same time, often across different levels of communication, a *contradiction* unfolds across time: an actor states or endorses one principle, and later behaves in a way that negates it (Nardone & Balbi, 2017). The distance between the first signal and the second makes the contradiction less immediately visible than a paradox, but not less powerful. In organizational life, contradictions often arise when people declare a desired direction and then repeatedly engage in behaviors that move them away from it.

A consultant may want more free time but continue to accept every new assignment. A manager may want a more autonomous team, but intervene every time a team member asks for help. In both cases, the contradiction is unintended and keeps the current pattern in place. Used intentionally, however, the same logic can become a lever for change. The same manager, for example, may provide the objective, the main constraints, and the criteria for evaluating a good decision, but then deliberately avoid following people step by step. This creates a productive tension. On the one hand, the manager provides direction and shows that the goal matters; on the other, they become less available as a constant source of reassurance. This contradiction can foster change because it prevents support from becoming substitution. Team members must now interpret the direction, make choices, tolerate mistakes within acceptable bounds, and learn from the consequences. In this way, the manager's partial unavailability becomes a developmental device: it protects the space in which autonomy and learning can emerge.

[9] In organizational settings, facilitators should also account for power asymmetries and avoid using such experiments to secure compliance or override legitimate dissent.

### 6.1. Emotions as Gateways to Change

Emotions are central to the Awaken phase. Identifying the emotion underlying a problem or performance block helps determine which attempted solutions are likely to be in play, which mechanisms sustain the problem, and which strategies may be worth trying. Four emotions warrant particular attention here: *anger*, *fear*, *pain*, and *pleasure*. This grouping is a practitioner's heuristic rather than a taxonomy of affect, and it does not correspond to other classifications used in emotion research[10]. It is taken from a practice-based account of strategic coaching (Milanese & Mordazzi, 2014) and I retain it because each of the four is associated with a characteristic family of attempted solutions, which is the link the Expose phase requires. Anger can narrow perception and trigger impulsive reactions. Fear can inhibit action or lead to avoidance and excessive help-seeking, thereby reducing self-confidence. Pain can lead to withdrawal, protective narratives, or dependency on consolation. Pleasure can become absorbing when it provides immediate reward and delayed cost (Nardone, 2019). The aim is not to eliminate emotions, but to prevent them from automatically selecting behavior. Emotions can also coexist or mask one another; work then starts from the most intense, and the same problem may call for different interventions depending on the emotion that sustains it[11].

Anger often appears when people feel mocked, attacked, treated unfairly, or denied recognition. The most common attempted solutions include repression, escalation, and displacement. Repression may contain the emotion temporarily, but often makes it reappear elsewhere; escalation can damage credibility and relationships; displacement can harm people who are not responsible for the original trigger. Possible strategies include creating a bounded, safe channel for anger, translating the emotion into language through writing or structured reflection, and reflecting on what the angry response produces in the relationships. In leadership, for instance, a manager who reacts aggressively during a meeting may be asked whether the outburst communicates authority or weakness. Expressive writing and other structured forms of emotional processing can help translate emotion into language and reduce

[10] Emotion research offers several competing taxonomies, including Ekman's (1992) basic emotions and Plutchik's (1980) Wheel of Emotions.

[11] The strategies described below are illustrative and apply to circumscribed emotional reactions that do not involve marked distress or impair the person's functioning; otherwise, referral remains the appropriate course.

its spread through the system (Baikie & Wilhelm, 2005; Nardone, 2019; Pennebaker & Smyth, 2016).

Fear calls for a different logic. Avoidance and repeated help-seeking are among its most common responses. Both can reduce fear in the moment while strengthening it over time, because the individual never experiences their own capacity to face the situation. Control attempts are another common response. For circumscribed, low-intensity fear, nonclinical strategies include graded practice with manageable versions of the feared situation and shifting attention. Publicly acknowledging manageable vulnerability and observing others face the same situation calmly may also help. For example, a professional apprehensive about public speaking may begin by acknowledging a minor difficulty at the start of a presentation, making the audience more receptive. The same professional could also draw on the Stoic practice of *premeditatio malorum*[12], briefly picturing beforehand what could go wrong so that the feared scenarios become familiar, without trying to intensify fear and stopping if apprehension increases. *Paradoxical intention* and deliberate fear evocation push this kind of controlled confrontation much further; they originate in clinical practice and require psychological training (Frankl, 1975; Nardone, 2000, 2014b).

Emotional pain may follow the loss of a role, rejection, or the collapse of a valued project. The main attempted solutions used by the person experiencing pain, or by those around them, include denial, distraction, excessive or premature consolation, overdependence on others, and protective narratives that avoid confronting the painful meaning of the event. Excessive consolation, in particular, can unintentionally communicate incapacity and foster dependence. These responses may provide temporary relief, but they can prevent the pain from being processed and keep it active in the background. Possible strategies include allowing time for this process, using writing or narrative reconstruction to organize the experience, and receiving a supportive presence without substitution. For example, after a professional failure, a coach or leader should not immediately minimize the event or provide ready-made solutions. A more useful response may be to help the person name what has been lost and identify what can be learned. Writing and narrative reconstruction can help people process difficult events by organizing them and allowing meaning to emerge (Pennebaker & Beall, 1986; Pennebaker & Smyth, 2016).

Pleasure maintains ineffective patterns when it becomes absorbing or costly. The most common responses, indulgence and rigid suppression, tend to reinforce the behavior or intensify the desire for it. Possible strategies include understanding the need beneath the behavior, replacing one pleasure with a more constructive one, containing the behavior within defined times and places, introducing friction or delay between impulse and action, progressively reducing low-value repetitions, and turning the behavior into a deliberate routine so that it becomes less automatic (Nardone, 2019). For example, a professional who frequently scrolls through social media during working hours may not need the platform itself, but rather relief from work and mental overload. Instead of simply suppressing the behavior, they could schedule a short walk, a conversation, or another pleasant activity that meets the same need in a more functional way. Similarly, limiting scrolling to fixed times and durations, such as five minutes at the beginning of each hour, can transform the behavior from an impulsive reaction into a planned action, reducing its immediate appeal. Reframing can support changes in how individuals perceive their behavior and help address the emotions discussed above. For example, a manager who stays late at the office every day may need help recognizing whether work has become their main source of pleasure, recognition, or escape.

[12] *Premeditatio malorum* is a Stoic practice of deliberately imagining possible difficulties, losses, or setbacks before they happen. Its goal is not pessimism, but preparation. The idea is that by imagining adversity in advance, one may become calmer and less surprised when difficulties arise.

The Awaken phase emphasizes experimentation. Once a possible strategy is identified, the actor tests it and observes its effects. A strategy that proves ineffective is diagnostic information, not mere failure. One can return to earlier phases to gather better data, redefine the goal, or identify an overlooked attempted solution.

The phase is complete when the actor has tested at least one strategy that modifies the maintaining pattern and has begun to experience a different relationship with the problem or objective.

## 7. Nurture

Nurture consolidates and integrates change into everyday life. A change that has been activated but not maintained remains fragile (Kwasnicka et al., 2016). A person can interrupt an ineffective habit and return to it under stress. A team can improve communication after a workshop and slide back to old routines as deadlines approach. This phase is about making new behaviors, perceptions, and relational patterns stable without making them rigid. Theories of behavior change maintenance point to motives, self-regulation, resources, habits, and environmental and social influences (Kwasnicka et al., 2016), all of which recur in this phase. The phase begins with structured reflection on what has changed and why.

Once a problem has been solved or a goal has been reached, the person involved should reconstruct what happened: which attempted solutions were in place, which actions interrupted them, which experiences modified perception, and which context and conditions supported the change. This reflection is fundamentally distinct from an initial search for causes in the past. Before action, many interpretations remain hypotheses; after successful action, the system has produced evidence. A useful question at this stage is *what could be done to undo everything achieved so far*, a formulation that applies the paradoxical logic introduced earlier to make relapse paths visible. Valuing what has been achieved before moving on also matters.

Consolidation also requires appropriate attribution. When a coach or leader accompanies someone through change, the credit for progress should remain with the person or group that enacted it. If progress is attributed mainly to the facilitator, autonomy remains weak. By contrast, if those involved recognize that change emerged from their own responses, actions, and choices, self-efficacy increases. A useful intervention should not create dependency. In coaching, this may mean closing within a set number of meetings and offering only a follow-up later. In leadership, it may mean progressively reducing guidance as the team learns how to act.

Repetition is the most direct mechanism of consolidation. Once an action interrupts a counterproductive pattern and increases performance, it should be repeated until it becomes a sustainable habit. Consistency often matters more than intensity, and commitments are more sustainable when they concern the process rather than the outcome, such as practicing for a fixed time each day rather than reaching a given result. Positive reinforcement is also important: when people perceive benefits from a new behavior, they are more likely to repeat it (Judah et al., 2018). Small improvements can accumulate to larger effects, which aligns with the logic of gradual consolidation. The relevant unit is not the heroic leap, but the repeated action. Over time, behavior can also become part of identity: the manager who repeatedly delegates begins to see themselves as someone who develops others rather than someone who must control everything.

Specificity supports habit formation. A new behavior should be connected to a clear time and place. It can also be attached to an existing routine, reducing decision effort. Yet habits should not become new constraints. A sleep routine that improves recovery is useful; a sleep rule that prevents all social life may become rigid. The person should persist in observing and moving in the direction of growth, rather than in mechanical repetition.

Setbacks should be treated as diagnostic information. *Implementation intentions* can also protect new behaviors from relapse by linking predictable high-risk situations to a predefined response in an if-then format (Gollwitzer & Sheeran, 2006). Klein's (2007) *premortem* technique can complement this planning process by asking individuals to imagine that the proposed plan has already failed and then identify plausible reasons for that failure, thereby revealing risks and possible preventive adjustments[13].

A further consideration concerns recovered time. When a problem loses power, it often releases attention previously consumed by rumination, conflict, avoidance, checking, excessive work, or ineffective attempted solutions. This recovered time is valuable but can be difficult to inhabit. Some individuals replace one problem with another because the absence of the old preoccupation leaves a void. Consolidation is stronger when recovered time nourishes the person, relationships, and broader personal projects rather than becoming another imbalance. A periodic review of life domains, such as the Wheel of Life (Whitworth et al., 1998), can help decide in advance how this time will be used.

### 7.1. Social Bonds, Self-Regulation, and Optimal Engagement

People develop through connections with others, their communities, and themselves; positive change is more likely to endure when all three forms of connection are in place. Positive social contact can support emotional regulation (Field, 2010), and social support is associated with lower stress and better well-being (Uchino, 2006). Strong social bonds provide safety, recognition, and support; weak ties, on the other hand, can provide novelty, nonredundant information, and opportunity (Granovetter, 1973). In change processes, connection offers feedback, encouragement, accountability, and access to resources that the individual does not have on their own.

[13] The premortem belongs to consolidation rather than to earlier phases because, while a goal is still being constructed, visualizing its attainment is more useful and motivating; once progress exists, imagining failure helps protect it.

However, connection with others requires boundaries and selectivity. It should not entail unconditional availability. Some relationships nourish growth; others repeatedly pull the individual toward conflict or dependency. Moreover, productive relationships can accommodate disagreement. An innovator, for example, may benefit from a partner who understands their perspective while challenging it without contempt (Verganti, 2017).

Connection extends beyond dyads to communities and environments. People tend to adopt behaviors that are normal in their groups. They are more likely to connect with similar others (McPherson et al., 2001), and shared identity can increase affiliation and influence (Cialdini, 2016). This has practical implications: someone who wants to maintain a new behavior is supported by contexts where that behavior is normal.

Connection also concerns the self. Frankl (1984) argued that meaning is central to human life. Accordingly, pursuing multiple meaningful personal projects provides continuity beyond short-term motivation and protects against over-reliance on any single source of identity or fulfillment. This plurality also shapes how setbacks are experienced: when identity rests on several domains, a single failure is less likely to be read as a verdict on the whole person. Individuals should learn to observe themselves without contempt. Excessive self-judgment can become another attempted solution: by attacking oneself, the person may believe they are enforcing improvement, while often increasing fear or avoidance. Failure should be embraced as part of learning, especially in innovation. The critical issue is whether failure becomes shame and withdrawal, or information and persistence.

Connecting with oneself also entails acceptance and presence. It involves recognizing and learning to manage emotions (Goleman, 2020) and becoming aware of what cannot be changed. Acceptance is not intended here as resignation. It means distinguishing what belongs to the past, what depends on others, and what can be acted upon now. Presence, in this sense, refers to the capacity to observe one's own thoughts and emotions as passing mental events without immediately identifying with or acting on them – a capacity grounded in the metacognitive tradition (Flavell, 1979) and central to acceptance-based approaches in behavioral psychology (Hayes et al., 1999). This metacognitive capacity, progressively refined through practice, is closely related to what Goleman (2020) identified as *emotional intelligence*: a cluster of competencies – including self-awareness and self-regulation – that significantly shapes individual performance, decision-making, and relational quality.

When self-awareness and self-regulation deepen alongside genuine engagement in meaningful activity, the conditions for "flow" become possible. Csikszentmihalyi (2008) describes flow as a state in which a person becomes fully absorbed in an activity that is challenging, intrinsically rewarding, and sufficiently aligned with their abilities. In this state, attention narrows without becoming rigid, distractions lose relevance, the perception of time may change, and action seems to proceed with less conscious effort. Flow often emerges after repeated practice, when skills have become sufficiently internalized, and the person no longer needs to control every movement, word, or decision consciously.

Flow cannot be scheduled or expected at every moment. It can, however, serve as a useful compass for personal development. When a person repeatedly experiences energy, concentration, and satisfaction in certain activities, roles, or contexts, those experiences may indicate a direction in which growth is more naturally sustained. It would be an error to think that successful personal development requires permanent exceptional performance at every moment. It requires the gradual creation of conditions in which effort, meaning, and capability become better aligned with the desired change. Sometimes the right move is subtractive: removing unnecessary control, bureaucracy, rumination, conflict, obligations, or relationships that no longer support growth. Sometimes it is additive: introducing play, recovery, social connection, or meaning where life has become too narrow.

Nurture is complete when change is no longer only an episode but part of how the individual lives and works. What worked has been reflected on, useful behaviors have been repeated, progress has been protected from relapse, supportive connections have been strengthened, boundaries have been set, and the conditions for growth have been identified and cultivated. The result is not final perfection. It is a more adaptive equilibrium, capable of supporting further improvement.

## 8. Discussion and Conclusions

This paper introduces a five-phase conceptual framework for meaningful change. Observe develops a systemic and present-oriented understanding of the situation. Create turns that understanding into a meaningful, actionable direction. Expose identifies the attempted solutions, resistances, biases, and self-deceptions that keep the current pattern in place. Awaken activates new behaviors and experiences through planning, reframing, paradoxical movement, and emotion-based experimentation. Nurture consolidates what works through habits, relationships, identity, and sustainable contexts. The framework can help coaches, leaders, and HR professionals avoid premature intervention, define goals that are owned by the people they support, identify patterns that sustain resistance, and design more effective developmental processes. Appendix B brings the OCEAN phases and their key elements together in a single table, providing a practice-oriented reference.

The primary contribution is conceptual and addresses the process models that practitioners use to structure developmental conversations in coaching and adjacent nonclinical settings. Within that literature, OCEAN proposes a different phase architecture, one that places systemic observation before goal construction and adds a dedicated phase for the mechanisms that maintain the current pattern. In doing so, it incorporates into a general process model the argument that a persistent difficulty is often sustained by the repeated attempt to remove it, an argument that strategic approaches to coaching and problem-solving have already applied outside clinical settings (Milanese & Mordazzi, 2014; Nardone, 2009). The framework draws on systemic and interactional thinking, constructivist assumptions about perceived reality, goal construction, the analysis of attempted solutions, and emotional awareness. Taken individually, many of the components OCEAN builds on are well established; its originality lies in how it orders and connects them, making explicit how problem perception, goal construction, maintaining mechanisms, behavioral activation, and consolidation depend on one another. Its contribution is thus not simply procedural: it highlights how neglecting the dependencies among these elements may explain the persistence or recurrence of ineffective patterns, even when individual interventions appear appropriate.

Without aiming to replace existing theories of change, coaching, leadership, or problem-solving, the framework also has implications for the domains in which it is applied. For work on change implementation, it supports reading resistance as information about the change process rather than as a deficit in those being asked to change (Ford & Ford, 2010). For leadership development, it specifies what a leader does in a developmental conversation, phase by phase, rather than which traits or styles are effective.

As already mentioned, OCEAN should also be distinguished from existing frameworks with an adjacent scope. For example, the GROW model structures coaching conversations around Goal, Reality, Options, and Will or Way Forward (Whitmore, 1992). While GROW typically begins by clarifying the goal, OCEAN places a dedicated observation phase before goal construction, because the first stated goal may itself reflect a partial reading of the situation, social pressure, defensive reasoning, or an ineffective attempted solution. Moreover, although the Reality phase in GROW explores the current situation and its obstacles, this model does not include a distinct and systematic phase equivalent to Expose, in which attempted solutions, resistances, biases, self-deceptions, and maintaining patterns are examined before new actions are designed. The same distinction applies to other approaches that overlap with OCEAN in their use of resources, exceptions, and future images (e.g., Cooperrider et al., 2008). The emphasis on uncovering the protective function of maintaining patterns finds its closest conceptual parallel outside this family of process models, in Kegan and Lahey's (2009) Immunity to Change. This approach shares OCEAN's premise that people may unknowingly engage in patterns that obstruct their stated goals. Its immunity map makes these patterns visible by surfacing hidden competing commitments and the assumptions that sustain them. Here too, however, the map begins from an improvement goal, whereas OCEAN places observation before goal setting.

While OCEAN draws selectively on solution-focused and strategic clinical traditions (e.g., De Shazer, 1988; Watzlawick et al., 1974; Watzlawick & Weakland, 1977), it is not a clinical protocol and does not diagnose or treat. It repositions selected elements of that logic within nonclinical contexts such as coaching and leadership. Its contribution goes beyond proposing a single intervention technique, such as a questioning method, a reframing exercise, or a paradoxical task. Rather, it organizes attentional, dialogic, systemic, emotional, and behavioral moves into a comprehensive change process. For the same reason, Transformative Dialogue should not be read as a standalone questioning method: although it may incorporate nondirective listening and solution-focused questioning, it is the communicative support on which all five phases rely.

A first limitation of OCEAN concerns its scope: the framework operates on the perceptions, meanings, emotions, and behaviors of individuals within the developmental conversation. It does not model the processes through which collective action is produced, such as coalition formation, the negotiation of goals among groups with divergent interests, or bounded rationality in organizational decision-making, resource allocation, or formal authority and power. For organization-wide change, it should therefore be used alongside approaches that address these processes, not as a substitute for them. This approach also does not work the same way for every person, group, or situation. Sometimes the system is not ready; sometimes the first intervention only lays the groundwork for subsequent change; sometimes another form of support is required. Indicators for referral include difficulties that generalize beyond the situation the work is focused on, persist outside it, or involve marked distress or impaired functioning.

The OCEAN framework has been iteratively refined through its use in individual and team coaching, and these experiences suggest practical promise. These experiences are not presented here as evidence of the framework's effectiveness. The primary purpose of this paper is to offer an initial conceptualization of OCEAN; systematic empirical evaluation will require dedicated research. Accordingly, Appendix A provides a brief illustration of the framework's application to a specific case rather than an empirical test. Future research should examine when the framework works best, for whom, and under which conditions – with case studies, longitudinal applications, intervention designs, and comparative analyses. Research could also test whether the phase structure helps practitioners avoid common errors, such as acting before observing, defining goals that are not owned, using more of the same attempted solution, or failing to consolidate early progress.

The OCEAN framework offers a structured way of thinking about change and remains open to both practical application and theoretical and empirical development. Future research could explore whether the framework, as a mode of inquiry rather than a narrow professional technique, can also be applied beyond developmental work, for example to innovation, strategy, marketing, and consumer understanding. I also plan to elaborate on the tools and exercises that can be associated with each phase. Another promising direction concerns measurement. Because the framework can be applied to different contexts, evaluation should not rely on a single metric. Coaching and developmental outcomes may include goal attainment, perceived self-efficacy and well-being, or changes in decision-making. Were the framework extended to collective settings, evaluation would also need to address indicators such as engagement, team learning, or conflict, together with the organizational processes that produce them. Future research could develop phase-specific indicators and also employ digital tools, such as AI and behavioral analytics, to provide mirrors of communication and interaction patterns that are difficult to observe directly (Fronzetti Colladon & Grippa, 2025). Used ethically and transparently, such tools may strengthen several phases by making hidden patterns visible and providing feedback over time.

### Declaration on the Use of Artificial Intelligence

During the preparation of this work, the author used ChatGPT from OpenAI and Claude from Anthropic to improve clarity, readability, and language quality, and to assist in summarizing longer conceptualizations originally developed by the author. After using these tools, the author reviewed and edited the content as needed and takes full responsibility for the content of the published article.

## Appendix A

### An Illustrative Application of the OCEAN Framework: Martha's Case

This Appendix briefly presents Martha's[14] case through the five phases of the OCEAN framework. Martha met with a coach to address what initially appeared to be a problem with time management and daily organization, particularly her difficulty arriving at work on time and starting the day effectively.

#### *Observe*

Martha contacted the coach because she was experiencing difficulties at work. At the beginning, she described her problem as an inability to organize her daily routines. She reported arriving late at work, preparing very slowly in the morning, and often failing to hear her alarm clock.

Through Transformative Dialogue, the coach gradually helped Martha reconstruct the context in which the difficulty appeared. Martha had recently changed her job, and her new role as European sales force coordinator at a large company had significantly changed the nature of her work. Unlike her previous position, this role required her not only to coordinate activities but also to speak in public and present results to sales teams from different countries. As the conversation became more detailed, a clearer pattern emerged: the days on which Martha arrived late at work were usually those that preceded business trips with her manager, when she knew that she would soon have to meet local teams in another country and present in English. On those evenings, she often stayed up late preparing, which made it harder for her to wake up the next morning.

This was significant because Martha did not feel confident in English and had never liked public speaking. Her manager's presence made the situation even more intimidating. He now lived in Italy, but was from the UK. When he was ill or unable to travel, Martha's presentations generally went better. Similarly, when she presented in Italy, she still felt some nervousness, but the difficulty was less intense. The problem, therefore, did not appear to concern her work performance in general. On the contrary, Martha performed very well, had good relationships with colleagues, and was appreciated by her manager. The moments when she struggled most were those when she had to present, especially in English and in the presence of her manager.

The coach explored what the problem was, where it occurred, with whom it occurred, and when it appeared. He discovered that Martha could still go on stage, wear the microphone, and deliver her presentations to hundreds of colleagues. During these presentations, however, she reported that her voice often sounded uncertain and that at times she lost her thread and paused for some time. She experienced the moments before going on stage quite negatively. At times, she wrote the most important points on her forearm, but even then, she sometimes forgot parts of the presentation. After presenting, even when it had gone well, she spent a long time going over the imperfections in her performance.

[14] The name "Martha" is fictional, as are some identifying details. The case is presented for illustrative purposes only and to show how the five phases can be used to structure a change process. It is not intended to establish a causal effect of the OCEAN framework or to provide generalizable evidence of effectiveness.

As a result, Martha and the coach developed a more precise understanding of the problem. It was not mainly a problem of morning disorganization. It was a problem connected to public speaking in the presence of her manager, especially when she had to speak in English.

Before agreeing on the goals of the work, the coach also checked that the difficulty was circumscribed: Martha's tension concerned presentations, especially in English and in front of her manager, and did not extend to other social or professional situations; she reported no signs of intense or persistent distress[15].

***Create***

Throughout their meetings, the coach continued to use Transformative Dialogue to help Martha reflect on her objective and the new conditions of her work. Her new role was no longer only office-based, as her previous position had been. It required her to serve as a motivational presence for the sales force and present European- and country-level results to local teams. In this sense, public speaking was not a marginal task, but an important part of her new professional identity.

When the coach used the miracle question, Martha did not answer by saying that she would hear the alarm clock and move efficiently and punctually to work. Instead, she began to imagine herself going to work happily and without apprehension, facing her days with a smile, and feeling appreciated by others even when she had to present in English. She described a situation in which she would live those moments with excitement and perform at her best. During that session, Martha felt, for a moment, as if she were already there, on stage, presenting successfully. That image gave her confidence and left her feeling energized for the rest of the day.

The goal that emerged was clear. Martha did not simply need to become more organized in the morning. She wanted to manage her nervousness when speaking in public and perform at her best when presenting in English while also being observed by her manager. This was a completely different objective from the one she had initially brought to coaching.

***Expose***

During the Expose phase, the coach asked Martha whether there were any exceptions and what she had already tried to do to solve the problem. This allowed her attempted solutions to emerge. Martha realized that at times she had asked her manager to present in her place, justifying the request by saying she wanted to learn from him. On other occasions, she had asked him to review her slides and listen to her presentation the day before, so that she could know whether she was doing well. However, because of his many commitments, he was not always able to do so.

[15] They agreed that the work would focus on her effectiveness as a presenter and that, should the difficulty prove broader or more intense, referral to a psychologist would be the appropriate course.

She also began to recognize that some of the strategies she used to feel safer had become obstacles. Memorizing the presentation or writing the key points on her forearm did little to help her stay in control. On the contrary, when she did not remember a section by heart, or when she realized that she had skipped part of the outline written on her forearm, she lost her thread. Once that happened, her tension and worry about making a bad impression increased, and the quality of her presentation declined. Similarly, she often studied late at night to imagine all the possible questions she might be asked and to prepare answers in English. This gave her the impression that she was preparing herself, but it also left her arriving at the presentation already tired.

By exploring what could make the situation worse, Martha recognized that her attempts to avoid the presentation, seek help, and control every possible detail were maintaining and worsening the problem. She realized that she would not gain confidence by continuing to ask others to shield her from these situations, or by trying to control the presentation more and more rigidly.

Some exceptions were also identified. Martha presented more calmly when her manager was not present. She was also calmer when she had already presented the material to him the day before. However, these exceptions were not fully replicable at will, because her manager was not always available and could not always be absent. More importantly, they did not address Martha's deeper need to show him that she was capable, competent, and well-prepared.

***Awaken***

The coach and Martha discovered that the prevailing emotion underlying the problem was fear (in a mild and circumscribed form: apprehension about presenting). More specifically, Martha was apprehensive about speaking in public, performing poorly in front of her manager, and appearing unprepared while speaking English. This apprehension had led her toward avoidance, requests for help, and attempts to control every possible scenario and every element of the presentation. Using paraphrasing, reframing, and emotionally evocative language typical of Transformative Dialogue, the coach helped Martha understand that every time she asked for help or avoided a presentation, she was actually worsening the problem by reinforcing her sense of incapacity. On the one hand, she felt safe or protected because she had avoided the situation or because her slides had been corrected. On the other hand, the implicit message she sent herself was: "On your own, you would not be able to perform at your best".

Martha also discovered the paradoxical effects of her attempt to control everything. Memorizing the presentation or writing notes on her arm made her lose control more easily. By concentrating too much on what she had to say, she often lost the thread of the presentation. By trying to memorize answers to all the possible questions she might be asked in English, she slept very little and arrived at the presentation already tired. Reframing produced an important first shift: Martha began to see those attempted solutions as counterproductive and wanted to drop them.

As the dialogue continued, the coach and Martha explored possible experiments and strategies that could help her act differently. She remembered that on some occasions, when her presentation was incomplete because some data were missing, she had openly declared this before beginning. Paradoxically, this had made her feel calmer. She realized that this transparency made the audience more benevolent toward her and less judgmental. She later used a similar strategy when presenting in London in front of native English speakers, joking at the beginning about her Italian accent. She discovered that openly acknowledging that her English was not perfect made her feel more relaxed.

By analyzing the stories of well-known speakers, including politicians and TED Talk speakers, Martha and her coach identified additional experiments to test. The first was to carry a small stone in her pocket and hold it in her hand when necessary during the presentation. This helped Martha especially at the beginning, when she felt the tension most strongly. Instead of focusing on her own tension, she focused on the sensation of the stone in the palm of her hand. This helped her shift attention away from her nervousness. In a sense, she was beginning to overcome it by looking elsewhere.

As another experiment, Martha decided to stop memorizing the presentation and writing notes on herself. To her surprise, she noticed that her presentations went better, not worse. Giving up excessive control did not make her more exposed. It made her feel freer and more effective.

Finally, in the days before her presentations, Martha spent some time imagining that the entire presentation would go wrong in the worst possible way. She imagined making a terrible impression in English, forgetting what she had to say, being unable to answer questions because of missing data, and being interrupted during the talk by an unexpected event. During those times, she did not try to reassure herself that these outcomes would not happen or to plan how to prevent them. This experiment followed an unconventional logic. Paradoxically, imagining the worst gradually took the edge off her apprehension. It was as if she became more familiar with those scenarios, as happens when one watches a horror film many times and, eventually, the monster is no longer frightening. Acting in the way she least expected helped her improve. The coach proposed the exercise as a bounded, voluntary experiment, agreed in advance and reviewed at the following session, with the explicit understanding that Martha would stop it if it increased her unease instead of reducing it. At each session, they also checked whether her apprehension was decreasing or increasing[16].

***Nurture***

Martha and the coach noted that her improvement was associated with a process of discovery, reframing, and new experiences, such as focusing on the object she held in her hand. She also changed through new, unconventional actions and ways of thinking, such as considering how to worsen the problem, imagining a miraculous scenario in which the problem had already been solved, and picturing in advance what could go wrong.

[16] In Martha's case, it was a bounded, nonclinical exercise addressing mild apprehension about a professional performance situation, not treatment for an anxiety disorder; such exercises should be avoided, and the person referred, when fear is intense, disabling, or clinically significant.

This was very different from what a public speaking trainer had previously suggested to her: to attend his course, to think positively that everything would go well, and to study nonverbal and paraverbal communication in depth to achieve full control of all elements while she was on stage. Martha's change followed a different route. Her progress came through a shift in perspective and new experiences, not through another attempt at control.

Overall, Martha and the coach reflected on the change achieved, the attempted solutions that had maintained the problem, and what she could do to undo all the progress made.

After reporting more effective presentations and less apprehension, Martha also learned to manage her time better. She practiced presentations more and more and, at the same time, began to take pleasure in her progress. This encouraged her to enroll in an improvisational theater course and to improve her English through a new app she downloaded on her phone.

Throughout the process, the coach avoided prescribing. He asked questions so that he and Martha could discover together which behaviors were sustaining the problem and what might interrupt them. By the final phase, this allowed the coach to return full credit for the change to Martha herself. She now felt much more confident and no longer depended on either her manager or her coach to perform at her best.

The coach met Martha again after three months, and she reported that the improvements had been maintained. She continued to describe greater confidence in presentations and less dependence on her manager's reassurance.

## Appendix B

Table B1 offers a brief overview of the OCEAN phases and their key elements, while also serving as a practice-oriented reference.

| Framework element | Observe | Create | Expose | Awaken | Nurture |
|---|---|---|---|---|---|
| **Purpose** | Develop a systemic, detailed, and present-oriented understanding of the situation before intervening. | Translate observation into a meaningful, owned, actionable, and measurable direction for change. | Identify what maintains the gap between the current situation and the desired state. | Support change through new actions, reframings, and emotionally significant experiences. | Consolidate change so that new behaviors, meanings, and relational patterns become stable without becoming rigid. |
| **Some key questions** | What is happening now? Who is involved? Where and when does the difficulty appear? What patterns repeat? What meanings, emotions, rules, and contextual factors shape the situation? What other explanations could produce the same signals or results? How do others see the person or group in this situation? How would others describe the problem, and what would they do? What would another person probably not think? What advice would one give a friend facing the same situation? | What should change? Why does it matter? Whose goal is it? What would be different if the goal were reached? How will progress be recognized? | What has already been tried? What happens after each attempted solution? Which thoughts, communications, or actions maintain the problem? What resistances, biases, or self-deceptions are present? Were there any exceptions, and would they still work today? What would make the problem worse? | What small action could interrupt the maintaining pattern? Beyond the first plausible solution, what alternatives exist? Does the situation require planning, experimentation, or a more emotional experience? What emotion is behind the current behavior, and could it be masking another? What happened when the new strategy was tested? | What changed? What made the change possible? What could undo the progress? What situations are most likely to trigger a return to the old pattern? How can the useful behavior be repeated? What will fill the time previously absorbed by the problem? In which activities does time seem to fly, and what gives or drains energy? What relationships, routines, environments, and boundaries support continuation? |
| **Some key practices and tools** | Present-oriented inquiry; systemic mapping of actors, interactions, circular patterns, and explicit and implicit rules; self-observation; perspective-expanding inquiry; functional “how” questions; attention-training. | Goal prioritization and distinction among actionable or negotiable aims, conditions requiring acceptance, and expectations to release; SMART goal formulation; objective indicators and anchored 1-10 self-assessment scales; the Miracle Question; acting “as if” through small present behaviors. | Structured analysis of attempted solutions; mapping of self-reinforcing loops; mapping habits and certainties; resistance analysis; analysis of past exceptions; metacognitive questioning of biases and self-deception; the Worsening Question; Problem Reversal. | Selection between rational and experiential pathways; brainstorming, mind maps, and the Six Thinking Hats; backward planning; reframing and emotionally significant new experiences; metacommunication; small, agreed behavioral experiments, including paradoxical or contradiction-based experiments when low-risk and appropriate; emotion-matched strategies such as expressive writing, attention shifting, graded practice for mild, nonclinical fears, and narrative reconstruction; reflection and recalibration after action. | Structured reflection on what changed and appropriate attribution of progress; gradual reduction of facilitator guidance and follow-up; repetition, positive reinforcement, and linking behaviors to stable times, places, or existing routines; implementation intentions (“if-then” plans); premortem; relapse-path mapping and the relapse question; constructive use of recovered time; periodic review of life domains; strengthening supportive relationships, communities, and environments; boundary setting; flow as a developmental compass. |
| **Expected output** | A shared description of the current situation, including key actors, interactions, contexts, perceived realities, and recurring patterns. | A clear and shared vision, translated into a realistic and motivating objective, with observable indicators of progress. | A clear map of recurring attempted solutions, maintaining mechanisms, resistances, biases, useful exceptions, and patterns that need to be interrupted. | A tested modification of the maintaining pattern, accompanied by a different relationship with the problem or goal. | A more stable and adaptive equilibrium, supported by habits, relationships, and contexts that make relapse less likely. |
| **Possible risks and common errors** | Acting too quickly; confusing interpretation with observation; relying excessively on "why" questions; blaming individuals instead of mapping interactional loops; ignoring emotions, implicit rules, or context. | Defining vague goals; accepting imposed goals without ownership; pursuing utopian or unrealistic objectives; failing to distinguish goals from wishes, constraints, or conditions requiring acceptance. | Searching only for original causes; blaming actors; overlooking inaction as an attempted solution; treating awareness as sufficient; recommending an extreme, opposite behavior. | Assuming insight is enough; overusing rational explanation when a block is emotional; prescribing action plans instead of helping the actor define them; encouraging experiments that are too large, risky, or insufficiently owned by the actor; treating an ineffective test as failure rather than information. | Ending too early; attributing success to the facilitator; turning new habits into rigid rules; failing to support the constructive use of recovered time; neglecting relational and environmental supports. |
| **Transition criteria** | Move forward when the current pattern is sufficiently clear to define a meaningful goal. Return to Observe if later phases reveal missing actors, misunderstood emotions, unclear context, or incomplete information. | Move forward when the goal is sufficiently owned, specific, actionable, and measurable. Return to Create if resistance emerges because the goal is not felt as legitimate, is too broad, or does not reflect the actor's real priorities. | Move forward when the maintaining pattern and the first target for interruption are clear. Return to Observe or Create if the problem definition changes, if the goal no longer fits, or if attempted solutions cannot be identified with enough specificity. | Move forward when a new behavior or experience has produced constructive change. Return to Expose if the intervention does not affect the maintaining mechanism. Return to Create if the action reveals that the goal was unrealistic, unclear, or not owned. | The process can close or shift to follow-up when the actor can understand and see what changed, repeat what works, detect relapse signals, and maintain progress. Return to previous phases if consolidation fails, new problems arise, or further goals and developmental needs emerge. |

*Note.* Transformative Dialogue – including questioning, active listening, paraphrasing, repetition, reframing, and emotionally resonant language – supports all five phases. The "Some key practices and tools" row highlights additional practices and tools most closely associated with each phase. The questions, practices, and tools presented in the table are illustrative rather than exhaustive or prescriptive. They are not intended to be used all at once or in every case, but should be selected as needed, when useful and appropriate to the context, the person or group involved, and the objectives of the current phase. In every phase, the work should be suspended and the person referred to a qualified professional whenever difficulties extend beyond the focal situation, involve marked distress or impaired functioning, or otherwise fall outside the scope of nonclinical work.

**Table B1.** The OCEAN Framework: Phase Overview and Operational Outline.